\documentclass[10pt,twocolumn]{article}
\usepackage[a4paper,margin=16mm,columnsep=6mm]{geometry}
\usepackage[T1]{fontenc}
\usepackage{lmodern,amsmath,amssymb,graphicx,natbib,caption,hyperref}
\hypersetup{colorlinks=true,linkcolor=blue,citecolor=blue,urlcolor=blue}
\newcommand{\AAng}{\ensuremath{\mathrm{\mathring{A}}}}
\newcommand{\avg}[1]{\left\langle #1\right\rangle}

\title{Characterization and empirical correction of a spatially coherent spectral artifact in Gemini/GMOS-N IFU observations}

\author{Xiaoyang Chen\\ \small Frontier Research Institute for Interdisciplinary Sciences, Tohoku University,\\ \small 6-3 Aramaki, Aoba-ku, Sendai, Miyagi 980-8578, Japan\\ \small astro@xychen.me}

\date{2026 November 25}

\begin{document}
\twocolumn[\maketitle\begin{@twocolumnfalse}

\begin{abstract}
We report a recurring spectral artifact identified in observations obtained with the integral field unit (IFU) of the Gemini Multi-Object Spectrograph on the Gemini North telescope (GMOS-N). The artifact appears as a vertically aligned structure in reconstructed images of the IFU field of view, associated with a specific group of fibers and exhibiting a broad spectral feature around 8000~\AAng. Similar features were repeatedly identified in observations of different astronomical targets obtained between 2020 and 2022 using the R150 grating in the two-slit IFU mode. Taking advantage of the reproducibility of the artifact in the fiber and wavelength domains, we develop an empirical correction method in which the artifact is modeled as the product of a common spectral profile and a fiber-dependent amplitude distribution, with an overall normalization determined independently for each exposure. The artifact template is constructed from observations in which the affected fibers are largely separated from the astronomical source and subsequently applied to independent observations in which the artifact overlaps the source. The correction substantially suppresses the artificial spatial structure in the reconstructed images. Although the physical origin of the artifact remains unidentified, we present its observational characteristics and correction procedure to facilitate the identification and treatment of similar features by other GMOS-IFU users.
\end{abstract}

\vspace{5mm}
\end{@twocolumnfalse}]

\section{Introduction}\label{sec:intro}
Integral field spectroscopy provides spatially resolved spectral information on astronomical sources and is widely used to investigate their spatial, kinematic, and physical properties. Reliable measurements of faint extended emission, however, require careful treatment of instrumental features that may introduce artificial spatial structures or spectral signatures into the reconstructed data cubes.

The Gemini Multi-Object Spectrograph (GMOS) provides integral field spectroscopy through an array of lenslets coupled to optical fibers. In the two-slit IFU mode, spectra obtained through 1000 fibers sampling the science field are distributed along two pseudo-slits and recorded on a mosaic of three CCDs. The spatial distribution of the incident light is subsequently reconstructed using the correspondence between the extracted spectra and the spatial positions of individual fibers \citep{gemini}.

Instrumental features in GMOS-IFU data cubes have previously been discussed by \citet{menezes2019}, who reported low-frequency spectral structures appearing as stripes in reconstructed images and presented correction methods based on principal component analysis (PCA) tomography.

During the reduction of GMOS-N IFU observations of a sample of galaxies, we independently identified a recurring artifact exhibiting a broad spectral excess around 8000~\AAng\ and a localized spatial distribution associated with a specific group of fibers. The artifact is particularly concerning when it overlaps faint extended astronomical emission, for which the separation of instrumental and astrophysical signals may be difficult.

\begin{figure}[!t]
\centering
\includegraphics[width=\columnwidth]{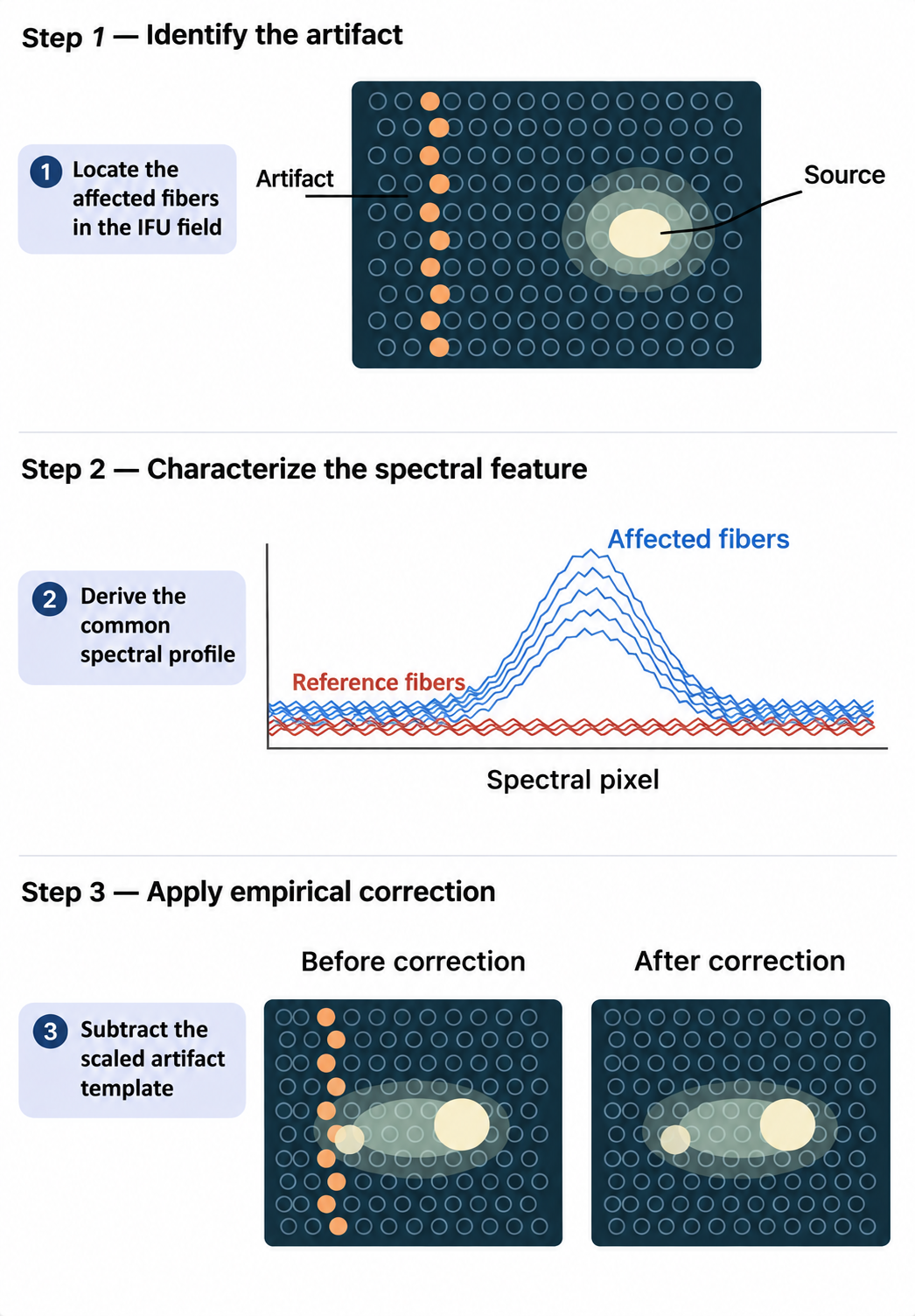}
\caption{Schematic illustration of the identification, characterization, and empirical correction of the spectral artifact in GMOS-N IFU observations. Step 1: identification of the artifact in the IFU field of view, where a vertically aligned artificial structure is spatially separated from the astronomical source. Step 2: characterization of the spectral profiles of the artifact using the affected and reference fibers. Step 3: application of the empirical artifact template to an independent observation in which the artifact partially overlaps the astronomical source. The schematic illustrates the principle of the correction and does not represent actual observational measurements.}
\label{fig:schematic}
\end{figure}

In this paper, we describe the observational characteristics of the artifact and present an empirical correction method exploiting its reproducibility across independent observations. Rather than investigating its physical origin, we aim to provide a practical reference for users encountering similar features in GMOS-IFU data. A schematic overview of the identification, characterization, and empirical correction of the artifact is presented in Figure~\ref{fig:schematic}.

\section{Observations and data reduction}\label{sec:obs}
The observations examined in this work were obtained with GMOS-N on the Gemini North telescope between 2020 and 2022 under programs GN-2020B-Q-117 (PI: X. Chen) and GN-2022A-Q-221 (PI: X. Chen).

The observations were carried out in the two-slit IFU mode, providing a science field of view of approximately $5''\times7''$ with a spatial sampling of $0.2''$. The R150 grating and GG455 filter were employed to obtain broad optical wavelength coverage. Different central wavelength settings were adopted during the observations.

The data were reduced using the Gemini IRAF package through PyRAF, supplemented by custom Python procedures. The reduction included bias subtraction, cosmic-ray rejection, bad-pixel treatment, scattered-light subtraction, spectral extraction, wavelength calibration, fiber-response correction, and spectrophotometric calibration. The extracted spectra were rectified onto a common wavelength grid extending from 4525 to 10804~\AAng\ with a sampling of 3.9~\AAng\ pixel$^{-1}$.

The artifact characterization and correction described below were performed on the extracted and wavelength-rectified fiber spectra after spectrophotometric calibration and prior to the reconstruction of the final science data cubes. Temporary data cubes without differential atmospheric refraction correction were constructed to examine the spatial distribution of the artifact in the original IFU coordinates.

For illustrative purposes, we present observations of two objects, hereafter referred to as J0941 (J094100.81+143614.5) and J0916 (J091624.75+073021.8). The observations of J0941 were used to construct the empirical artifact template, which was subsequently applied to the observations of J0916.

\section{Characteristics of the artifact}\label{sec:characteristics}
\subsection{Spatial distribution}
\begin{figure}[!t]
\centering
\includegraphics[width=\columnwidth]{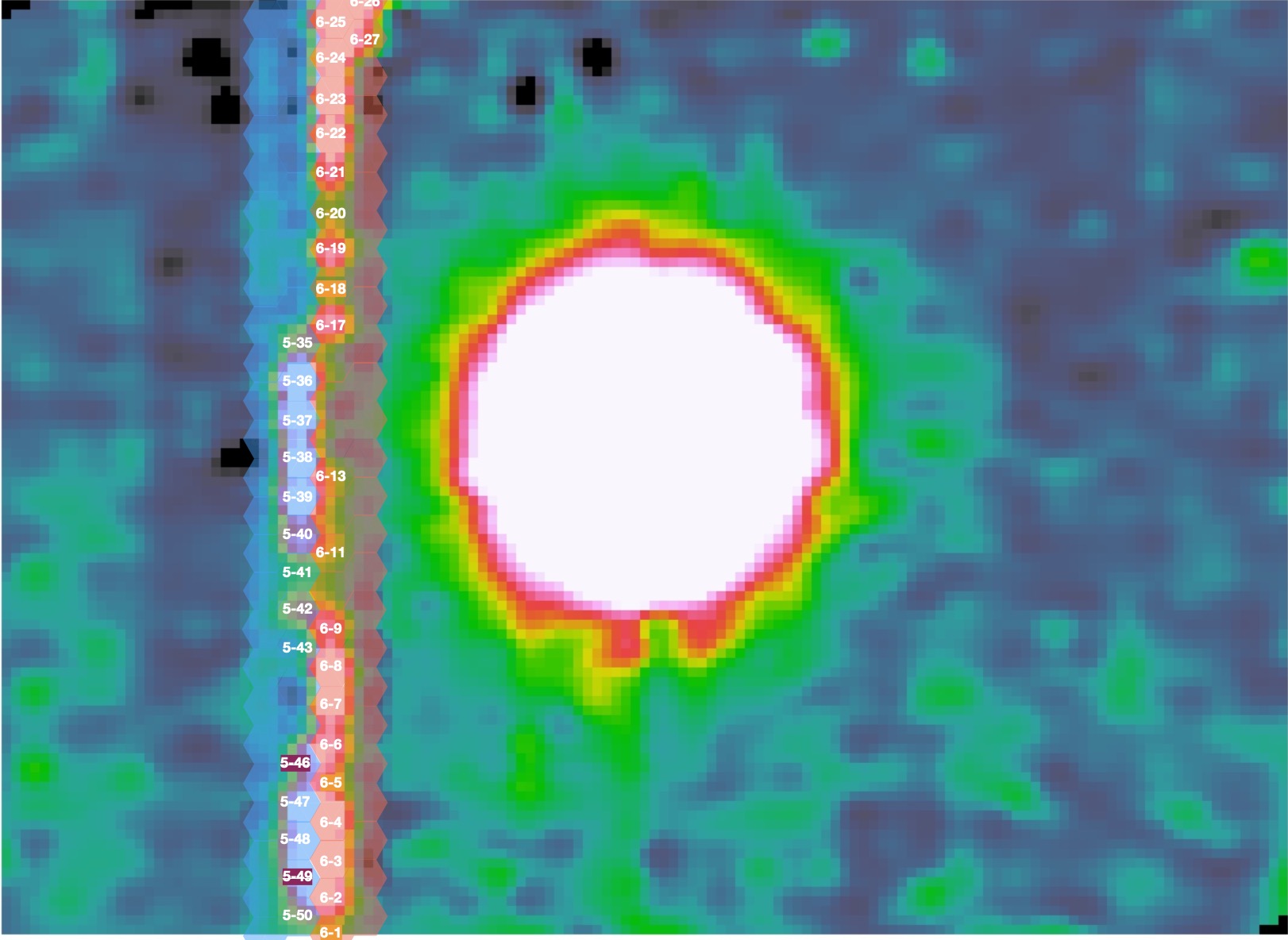}
\caption{Spatial distribution of the artifact in the GMOS-N IFU observations of J0941. The image is constructed over the wavelength region where the artifact is prominent. The artifact appears as a vertically aligned structure on the left-hand side of the field, largely separated from the astronomical source. The affected fibers are primarily associated with blocks 5 and 6. The hexagonal markers represent the spatial positions of the individual IFU fibers, and the labels indicate the identifiers of selected affected fibers.}
\label{fig:spatial}
\end{figure}

Figure~\ref{fig:spatial} shows a wavelength-stacked image of J0941 in the spectral region where the artifact is prominent. In addition to the emission associated with the astronomical source, an artificial, vertically aligned structure is apparent on the left-hand side of the IFU field. The artifact is associated with a group of fibers primarily distributed across fiber blocks 5 and 6.

A similar spatial structure is observed in J0916 (Figure~\ref{fig:correction}). Despite the different astronomical targets and observing epochs, the affected fibers are largely common to the two observations.

Similar features were encountered during the original reduction of observations of other targets obtained between 2020 and 2022. The recurrence of the feature at approximately the same fiber locations suggests that it is associated with the instrumental or data-reduction system rather than the intrinsic spatial distribution of individual astronomical sources. The spatial separation between the artifact and the main astronomical source in J0941 provides a favorable condition for empirical characterization of the artifact.

\subsection{Spectral characteristics}
\begin{figure}[!t]
\centering
\includegraphics[width=\columnwidth]{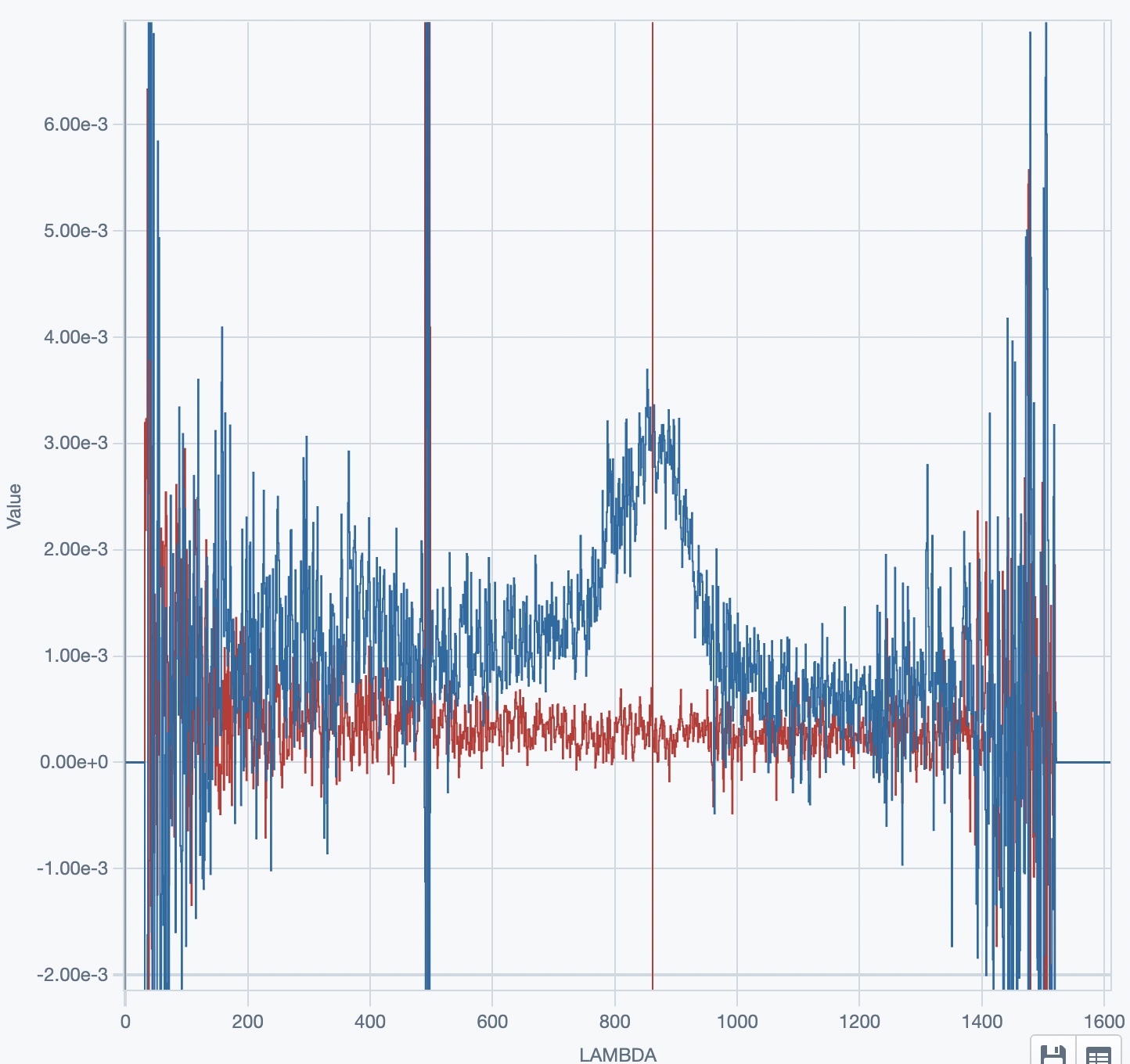}
\caption{Spectral characteristics of the artifact identified in the observations of J0941. Spectra of the affected fibers (blue) and reference fibers (red) are compared. The horizontal axis, labeled LAMBDA in the original plot, represents the spectral-pixel index in the wavelength-rectified spectra, with a wavelength sampling of 3.9~\AAng\ pixel$^{-1}$. The affected fibers exhibit a broad excess, particularly around spectral pixel 860, corresponding to approximately 7880~\AAng.}
\label{fig:spectral}
\end{figure}

The extracted spectra of fibers affected by the artifact are compared with those of reference fibers in Figure~\ref{fig:spectral}. The affected fibers exhibit an excess over a broad spectral range, with the most prominent feature located around 8000~\AAng. Although the amplitudes of the excess differ among individual fibers, their spectral profiles exhibit similar overall shapes.

The adopted correction region covers spectral pixels 720--1000 in the wavelength-rectified spectra, corresponding to approximately 7330--8420~\AAng. The strongest portion of the feature is located within approximately 7640--8110~\AAng, with a peak near 7880~\AAng. The similarities in the spatial distribution and spectral profiles among the observations motivate the empirical model described in the following section. We note that the selected wavelength intervals define the region adopted for the correction and do not necessarily represent precise physical boundaries of the artifact.

\section{Empirical correction method}\label{sec:method}
\subsection{Model formulation}
We model the observed spectrum of fiber $f$ in exposure $i$ as the sum of the astronomical signal, the background, and an additive artifact component:
\begin{equation}
D_i(f,\lambda)=S_i(f,\lambda)+B_i(f,\lambda)+M_i(f,\lambda)+\epsilon_i(f,\lambda),
\label{eq:data}
\end{equation}
where $S_i$ is the astronomical signal, $B_i$ is the background, $M_i$ is the artifact component, and $\epsilon_i$ represents observational noise.

Based on the observed similarities in the spectral profiles and spatial distributions of the artifact, we approximate the artifact component as
\begin{equation}
M_i(f,\lambda)=k_i A(f)P(\lambda),
\label{eq:model}
\end{equation}
where $A(f)$ represents the fiber-dependent amplitude distribution, $P(\lambda)$ is the common spectral profile, and $k_i$ is an exposure-dependent normalization factor. The product
\begin{equation}
T(f,\lambda)=A(f)P(\lambda)
\label{eq:template}
\end{equation}
defines an empirical artifact template in the fiber--wavelength domain. This formulation assumes that the relative distribution of the artifact among the fibers and its spectral shape are approximately reproducible across observations, while allowing its overall amplitude to vary between individual exposures (Figure~\ref{fig:schematic}).

\subsection{Construction of the artifact template}
The artifact template was constructed using the observations of J0941 obtained with a central wavelength setting of 720~nm. The observations obtained with the 680~nm setting were excluded because the CCD gaps compromised the spectral coverage of the artifact.

First, the affected fibers were identified by examining wavelength-stacked images constructed over spectral pixels 820--900, corresponding to approximately 7720--8030~\AAng. We selected 75 candidate apertures associated with fiber blocks 5 and 6 and adopted 100 apertures in blocks 2 and 3 as reference apertures largely unaffected by the artifact.

For each exposure, an average reference spectrum was constructed from the reference apertures and subtracted from the individual fiber spectra. The resulting residual spectrum is expressed as
\begin{equation}
R_i(f,\lambda)=D_i(f,\lambda)-\avg{D_i(f,\lambda)}_{f\in F_{\rm ref}},
\label{eq:residual}
\end{equation}
where $F_{\rm ref}$ is the set of reference apertures. The amplitude of the artifact in each fiber was estimated from the mean residual within the spectral interval where the artifact is most prominent (approximately 7640--8110~\AAng):
\begin{equation}
a_i(f)=\avg{R_i(f,\lambda)}_{\lambda\in W_{\rm strong}}.
\label{eq:amp}
\end{equation}

To estimate the common spectral profile independently of the fiber-dependent amplitudes, the residual spectrum of each fiber was normalized by its corresponding amplitude $a_i(f)$. The normalized residual spectra of 24 apertures exhibiting prominent artifact features were averaged, followed by averaging across the selected exposures.

The combined spectrum was subsequently fitted with a Chebyshev function using the IRAF task \texttt{fit1d}, with an order parameter of 36 and iterative asymmetric rejection of deviant spectral pixels. Bad-pixel information was incorporated into the fitting procedure. The resulting smooth spectral profile, $P(\lambda)$, was set to zero outside approximately 7330--8420~\AAng.

The spatial amplitude distribution was derived independently by calculating the mean and scatter of $a_i(f)$ across the selected exposures:
\begin{align}
\bar a(f)&=\avg{a_i(f)}_i,\label{eq:mean}\\
\sigma_a(f)&=\operatorname{std}_i[a_i(f)].\label{eq:scatter}
\end{align}
To avoid introducing a correction into fibers without reproducible evidence of contamination, we retained only candidate affected apertures for which $\bar a(f)$ exceeded twice $\sigma_a(f)$. The amplitude was set to zero for the remaining apertures. This is an empirical mean-to-scatter criterion rather than a formal detection significance. The final two-dimensional artifact template was constructed by multiplying the spatial amplitude distribution $A(f)$ by the common spectral profile $P(\lambda)$, as expressed in equation~(\ref{eq:template}).

\subsection{Application to individual observations}
The artifact template constructed from J0941 was subsequently applied to individual exposures of J0916. Because the absolute amplitude of the artifact may vary between exposures, the template was normalized independently for each exposure before subtraction.

As in the template construction procedure, the average spectrum of the unaffected reference apertures was subtracted from the extracted fiber spectra to obtain residual spectra. For each fiber, the mean residual within the full artifact wavelength interval was compared with the corresponding mean value of the artifact template:
\begin{equation}
r_i(f)=\frac{\avg{R_i(f,\lambda)}_{W_i(f)}}{\avg{T(f,\lambda)}_{W_i(f)}},
\label{eq:ratio}
\end{equation}
where $W_i(f)$ represents the available spectral pixels within approximately 7330--8420~\AAng. The same bad-pixel mask was applied to both the observed residual and the artifact template when calculating their wavelength averages, thereby accounting for differences in spectral coverage caused by CCD gaps and flagged pixels.

A subset of 12 candidate affected apertures with negligible expected galaxy contribution was adopted for estimating the exposure-dependent normalization. The normalization factor $k_i$ was calculated as the mean of the positive amplitude ratios among these apertures. The corrected fiber spectra were obtained as
\begin{equation}
D_{i,\mathrm{corr}}(f,\lambda)=D_i(f,\lambda)-k_i T(f,\lambda).
\label{eq:corrected}
\end{equation}
Note that the background-subtracted residual spectra were used only for estimating the artifact normalization. The scaled artifact template was subtracted from the original calibrated fiber spectra. The corrected spectra were then processed through the subsequent data-reduction procedures to reconstruct the final science data cubes.

An advantage of this approach is that the artifact contribution in fibers overlapping the astronomical source can be estimated without directly fitting the potentially contaminated astronomical spectra in those fibers.

\section{Correction results}\label{sec:results}
\begin{figure*}[!t]
\centering
\begin{minipage}[t]{0.48\textwidth}\centering
\textbf{Before correction}\\[2pt]
\includegraphics[width=\linewidth]{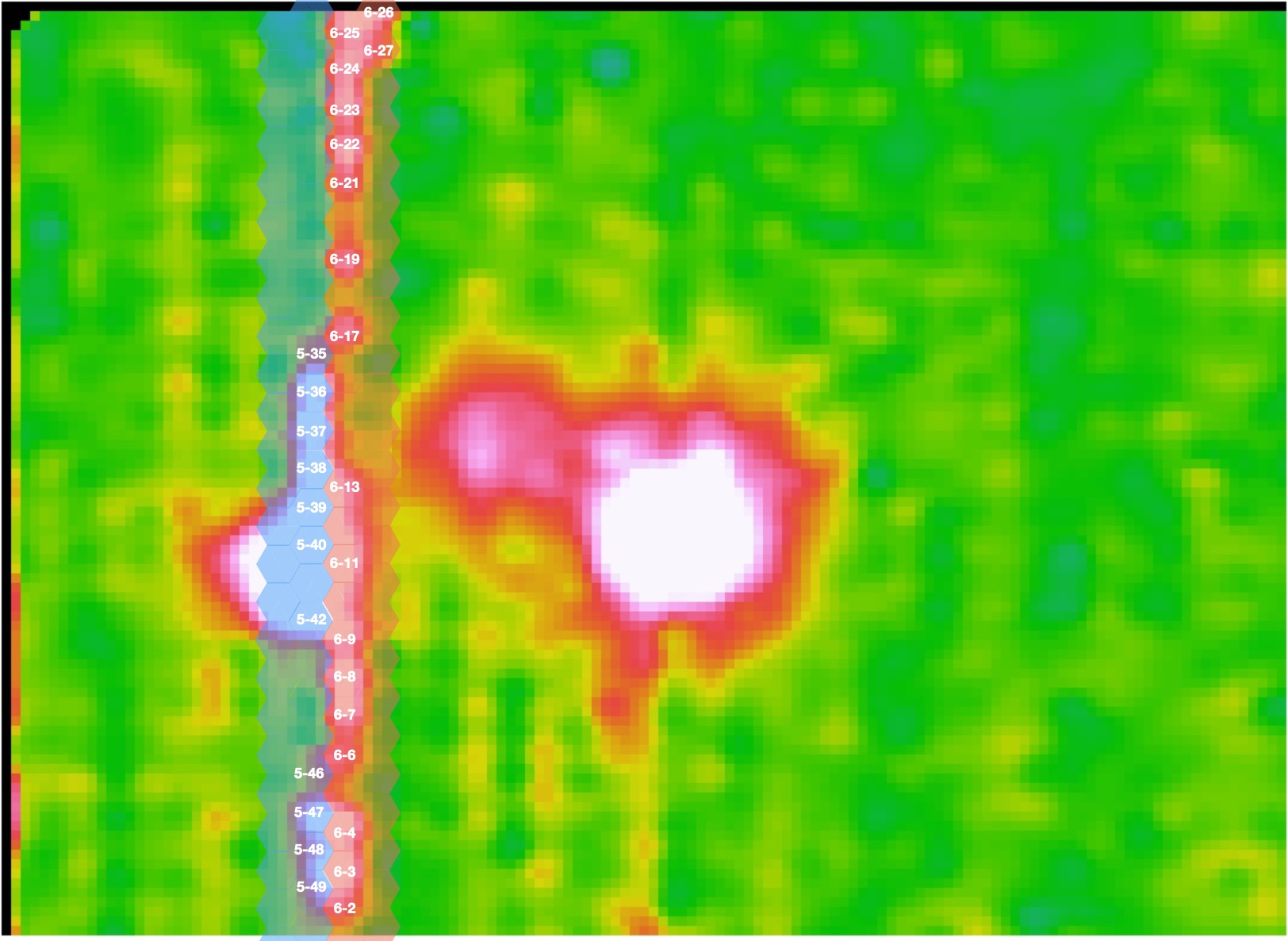}
\end{minipage}\hfill
\begin{minipage}[t]{0.48\textwidth}\centering
\textbf{After correction}\\[2pt]
\includegraphics[width=\linewidth]{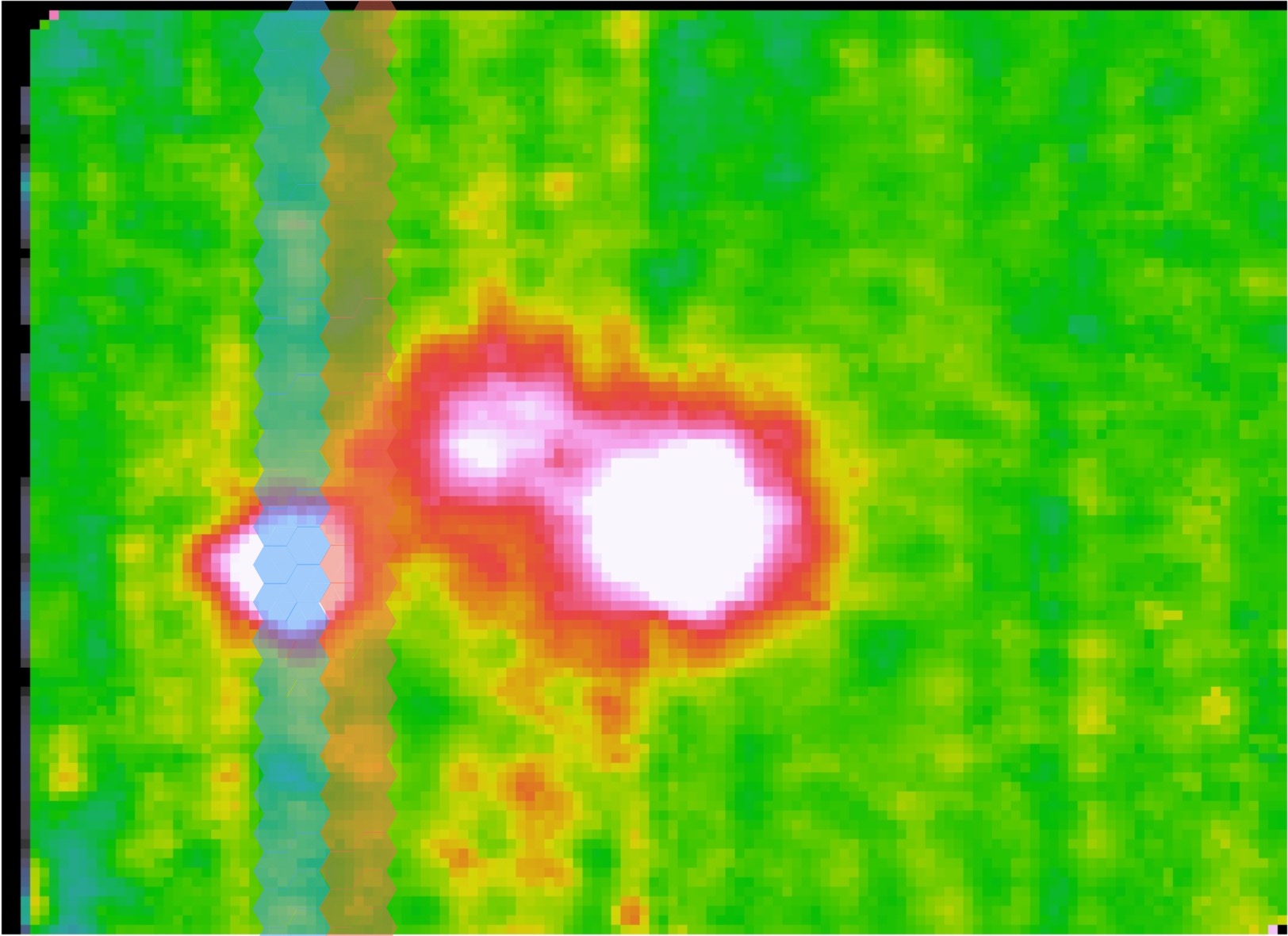}
\end{minipage}
\caption{Comparison of the spatial distribution of emission in the GMOS-N IFU observations of J0916 before (left) and after (right) the empirical artifact correction. The artifact template is constructed independently from the observations of J0941 and scaled for the individual exposures of J0916. The artificial structure on the left-hand side of the IFU field is substantially suppressed after the correction.}
\label{fig:correction}
\end{figure*}

Figure~\ref{fig:correction} compares wavelength-stacked images of J0916 before and after the artifact correction. Before correction, the vertically aligned artificial structure is clearly visible on the left-hand side of the IFU field. Unlike the case of J0941, the artifact partially overlaps the spatial distribution of the astronomical source, making a simple exclusion of the affected fibers undesirable.

After subtraction of the scaled artifact template, the artificial structure is substantially suppressed, while the main spatial structure associated with the astronomical source remains apparent. The suppression of the artifact using a template constructed from an independent astronomical target indicates that its spatial and spectral characteristics are sufficiently reproducible between these observations for the empirical correction to be effective. 
The correction has been employed in the data reduction for \citet{2025ApJ...979...32C} and \citet{2026ApJ...999...30C}.

The correction is particularly relevant for spatially resolved analyses of faint extended emission, for which contamination from the artifact may otherwise introduce artificial spatial structures or bias spectral measurements. The comparison presented here provides a qualitative demonstration of the correction. A quantitative assessment of the residual contamination and the recovery accuracy of the intrinsic astronomical signal is beyond the scope of this work.

\section{Discussion and limitations}\label{sec:discussion}
\subsection{Relation to previously reported instrumental features}
Spatially coherent instrumental features in GMOS-IFU data cubes have previously been reported by \citet{menezes2019}. Their instrumental fingerprints exhibit low-frequency spectral signatures and appear as stripes in reconstructed images. They proposed correction procedures based on PCA tomography, including its combination with wavelet decomposition.

The artifact presented in this work shares some observational characteristics with those previously reported features, particularly the striped spatial morphology and broad spectral signature. However, whether these phenomena have the same physical origin remains unclear. The correction presented here differs from the previous approaches in exploiting the correspondence between individual fibers and their artifact amplitudes, together with the reproducibility of the artifact across independent observations. The correction is performed directly on the extracted fiber spectra before the reconstruction of the final data cubes. The two approaches may therefore be regarded as complementary empirical methods for treating spatially coherent instrumental features.

\subsection{Applicability and limitations of the empirical correction}
The physical origin of the artifact has not been identified. Although its recurrence at similar fiber locations suggests an association with the instrumental system, the present investigation does not establish which instrumental component or data-reduction process is responsible for the observed feature.

The artifact was not apparent in the flat-field exposures examined during the original reduction. Whether this is related to the exposure duration or differences in the illumination conditions remains unclear.

Our correction assumes that the spectral profile and relative fiber-dependent amplitudes of the artifact are approximately stable across observations. Temporal or configuration-dependent variations in these quantities may limit the accuracy of the correction. The normalization procedure also assumes that a subset of affected fibers is available in which the astronomical contribution is negligible. For spatially extended sources filling the entire IFU field, the normalization may be less reliable or require an independent estimate.

Furthermore, subtraction of the empirical artifact template does not remove the observational noise associated with the original contamination. The uncertainty introduced by the construction and normalization of the template has not been independently quantified. The original implementation adjusted the variance array by adding the numerical artifact model during subtraction. This empirical treatment has not been validated as a statistically or dimensionally consistent propagation of uncertainty and should not be regarded as a rigorous variance prescription.

Finally, the observations examined in this work were obtained with a particular configuration of GMOS-N, employing the R150 grating in the two-slit IFU mode. The presence and characteristics of similar artifacts in observations obtained with other gratings, IFU configurations, detector systems, or at later epochs have not been systematically investigated. The template derived in this work should therefore be regarded as an empirical correction appropriate to the observations examined here, rather than a universal calibration applicable to all GMOS-IFU observations.

\section{Summary}\label{sec:summary}
We have presented the observational characteristics and an empirical correction of a recurring spectral artifact identified in GMOS-N IFU observations obtained between 2020 and 2022. The main results are summarized as follows.
\begin{enumerate}
\item The artifact exhibits a broad spectral excess around 8000~\AAng\ and appears as a vertically aligned artificial structure in reconstructed IFU images.
\item The artifact is associated with a specific group of fibers, primarily in fiber blocks 5 and 6, and similar spatial distributions are observed for different astronomical targets.
\item An empirical artifact template is constructed as the product of a common spectral profile and a fiber-dependent amplitude distribution. An overall normalization is independently determined for each exposure.
\item The application of the template derived from J0941 to independent observations of J0916 substantially suppresses the artificial spatial structure.
\end{enumerate}
Although the physical origin and broader applicability of the artifact correction remain to be investigated, the reproducibility of the feature provides a practical means of treating the contamination in the observations examined here. We present these results to assist other GMOS-IFU users in identifying similar instrumental features and mitigating their potential effects on scientific analyses.

\section*{Data availability}
The observational data used in this work are available through the Gemini Observatory Archive (\url{https://archive.gemini.edu/}) under program IDs GN-2020B-Q-117 and GN-2022A-Q-221.
The empirical artifact template and correction scripts are available from the author upon request (astro@xychen.me).

\section*{Acknowledgements}
We thank Masayuki Akiyama for fruitful discussions on optical instrumentation and instrumental effects.
We also thank Chia (a generative AI assistant) for fruitful discussions and assistance in the preparation of this manuscript, including the schematic illustration in Figure 1.
This work is supported by the Japan Society for the Promotion of Science (JSPS) KAKENHI Grant Number 25K17447.
The observations were carried out within the framework of Subaru-Gemini time exchange program which is operated by the National Astronomical Observatory of Japan. We are honored and grateful for the opportunity of observing the Universe from Maunakea, which has the cultural, historical and natural significance in Hawaii.

\end{document}